\documentclass[aps,prd,twocolumn,preprintnumbers,showpacs,nofootinbib,amssymb]{revtex4}
\usepackage{graphicx}
\usepackage{amsmath}
\usepackage{amssymb}
\usepackage{amsfonts}
\usepackage{bm}
\usepackage{color}

\def\be{\begin{equation}}
\def\ee{\end{equation}}
\def\bea{\begin{eqnarray}}
\def\eea{\end{eqnarray}}

\begin{document}

\title{The Logotropic Dark Fluid as a unification of dark matter and dark
energy}
\author{Pierre-Henri Chavanis}
\affiliation{Laboratoire de Physique Th\'eorique,
Universit\'e Paul Sabatier, 118 route de Narbonne  31062 Toulouse, France}

\begin{abstract}  We propose a heuristic unification of dark
matter and dark
energy in terms of a single ``dark fluid'' with a logotropic equation of
state $P=A\ln(\rho/\rho_P)$, where $\rho$ is the rest-mass
density, $\rho_P=5.16\times 10^{99}\, {\rm g}\, {\rm m}^{-3}$ is
the Planck density, and $A$ is the logotropic temperature. The energy density
$\epsilon$ is the sum of a rest-mass energy term $\rho c^2$ mimicking dark
matter and an internal energy term $u(\rho)=-P(\rho)-A$ mimicking dark energy. 
The
logotropic
temperature is approximately given by $A \simeq
\rho_{\Lambda}c^2/\ln(\rho_P/\rho_{\Lambda})\simeq\rho_{\Lambda}c^2/[123
\ln(10)]$, where
$\rho_{\Lambda}=6.72\times 10^{-24}\, {\rm g}\, {\rm m}^{-3}$ is the
cosmological density and $123$ is the famous number appearing in the
ratio $\rho_P/\rho_{\Lambda}\sim 10^{123}$ between the Planck density  and the
cosmological density. More precisely, we obtain $A=2.13\times 10^{-9}
\, {\rm g}\, {\rm m}^{-1}\, {\rm s}^{-2}$ that we interpret as a fundamental
constant. At the cosmological
scale, this model fullfills the
same observational constraints as the $\Lambda$CDM model (they
will differ in about $25$ Gyrs when the logotropic universe becomes phantom).
However, the logotropic dark fluid has a
nonzero speed of sound and a nonzero Jeans length which, at the beginning of
the matter era, is about $\lambda_J=40.4\, {\rm pc}$, in agreement
with the minimum size of the dark matter halos observed in the universe. At the
galactic scale, the logotropic pressure balances gravitational attraction and
solves the cusp problem and the missing satellite problem. The logotropic
equation of state generates a universal rotation curve that
agrees with the empirical Burkert profile of dark matter halos up to the halo
radius. In addition, it implies that all the dark
matter halos have the same surface density  $\Sigma_0=\rho_0 r_h=141\,
M_{\odot}/{\rm
pc}^2$ and that the mass of dwarf galaxies enclosed within a sphere of fixed
radius
$r_{u}=300\, {\rm pc}$ has the same value
$M_{300}=1.93\times 10^{7}\, M_{\odot}$, in remarkable agreement with the
observations. It also implies the Tully-Fisher relation $M_b/v_h^4=44
M_{\odot}{\rm km}^{-4}{\rm s}^4$.  We stress that there is no free parameter in
our model (we predict the  values of $\Sigma_0$, $M_{300}$ and   $M_b/v_h^4$
in terms of fundamental constants). We sketch
 a justification of the logotropic equation of state in
relation to  the
Cardassian model (motivated by the existence of extra-dimensions) and in
relation to Tsallis generalized thermodynamics. We also
develop
a scalar
field theory based on a Gross-Pitaevskii equation with an inverted quadratic
potential, or on a Klein-Gordon equation with a logarithmic potential.
\end{abstract}


\maketitle

\section{Introduction}
\label{sec_intro}

The nature of dark matter
(DM) and dark energy (DE) is still unknown and remains one of the greatest
mysteries of modern cosmology. DM has been introduced in
astrophysics to
account for the missing mass of the galaxies inferred from the virial theorem
\cite{zwicky} and to explain their flat rotation curves 
\cite{flat}. DE has been introduced in cosmology to account for the
present
acceleration of the expansion of the universe \cite{novae}. In the standard
cold dark matter ($\Lambda$CDM) model, DM is represented by a pressureless fluid
and DE is ascribed to the
cosmological constant $\Lambda$ introduced by Einstein
\cite{einsteincosmo}.
The $\Lambda$CDM model
works remarkably well at the cosmological scale but it encounters serious
problems at the galactic scale. In particular, it predicts that DM halos
should be cuspy \cite{nfw} while observations reveal that they have a flat 
core \cite{observations}. On the other hand, the $\Lambda$CDM model predicts
an over-abundance of small-scale structures (subhalos/satellites), much more
than what is observed around the Milky Way \cite{satellites}. These problems are
referred to as the ``cusp problem'' and  ``missing satellite problem''. The
expression ``small-scale crisis of CDM'' has been coined. 

There are
also unexplained important observational results. 
For example, it is an empirical fact that the surface density of galaxies has
the same value $\Sigma_0=\rho_0 r_h=141_{-52}^{+83}\, M_{\odot}/{\rm pc}^2$ even
if
their sizes and masses vary by several orders of magnitude (up to $14$
orders of magnitude in luminosity)
\cite{donato}. On the other hand, it is known that the
asymptotic circular velocity of the galaxies is related to their baryonic mass
by the
Tully-Fisher (TF) relation $M_b/v_h^4=47\pm 6 M_{\odot}{\rm km}^{-4}{\rm s}^4$
\cite{tf,mcgaugh}. Finally, Strigari {\it et al.} \cite{strigari} have shown 
that
all  dwarf
spheroidal galaxies (dSphs)
of the Milky Way have the same total DM mass contained within a radius
of $r_u=300\, {\rm pc}$. From the observations, they obtained $\log
(M_{300}/M_{\odot})=7.0^{+0.3}_{-0.4}$. To our knowledge, there is no
theoretical explanation of these observational results.

The small scale problems of the $\Lambda$CDM model are related to the
assumption that DM is pressureless. This assumption is valid if
DM is made of weakly interacting massive particles (WIMPs) with a mass in the
GeV-TeV range. These particles freeze out from thermal equilibrium in the
early universe and, as a consequence of this decoupling,  cool off rapidly as
the universe expands. In order to solve the small-scale crisis of
CDM, some authors have developed alternative models of
DM. For example, it has been proposed that DM halos are made of fermions (such
as sterile neutrinos) with a mass in the keV
range \cite{vega,clm}, or bosons (such as axions)
in the form of Bose-Einstein
condensates
(BECs) with a mass ranging from $10^{-2}\, {\rm eV}$ to $10^{-20}\, {\rm eV}$
depending whether the bosons interact or not \cite{bosons}.
In these models, the quantum pressure prevents gravitational collapse and leads
to cores instead of cusps. These models sometimes provide a good fit of
the rotation curves of galaxies but they do  not explain the universality (and
the values) of $\Sigma_0$, $M_b/v_h^4$, and $M_{300}$.

On the other hand, at the cosmological scale, despite its success at
explaining many observations, the $\Lambda$CDM model has to face two theoretical
problems. The first one is the cosmic
coincidence problem, namely why the ratio of DE and DM is of order
unity today if they are two different entities \cite{ccp}. The second one is the
cosmological constant problem \cite{weinbergcosmo}. The
cosmological constant $\Lambda$
is equivalent to a constant energy density
$\epsilon_{\Lambda}=\rho_{\Lambda}c^2=\Lambda c^2/8\pi
G$ associated with an equation of state $P=-\epsilon$
involving a negative
pressure. Some authors \cite{vacuum} have proposed to
interpret
the  cosmological constant in terms of the vacuum energy.
Cosmological
observations lead to the value
$\rho_{\Lambda}=\Lambda/8\pi G=6.72\times 10^{-24}\,
{\rm g}\, {\rm m}^{-3}$ of the cosmological density (DE).  However, particle
physics and quantum field theory
predict that the vacuum energy should be of the order of the Planck
density $\rho_P=c^5/\hbar G^2=5.16\times 10^{99}\, {\rm g}\, {\rm m}^{-3}$. The
ratio between
the Planck density $\rho_P$ and the cosmological density $\rho_{\Lambda}$ is
\begin{equation}
\frac{\rho_P}{\rho_{\Lambda}}\sim 10^{123},
\label{l1}
\end{equation}
so these quantities differ by $123$ orders of magnitude! This is the origin of
the cosmological constant problem. To circumvent this
problem, some authors have proposed to abandon the cosmological constant
$\Lambda$ and to explain the acceleration of the universe in terms of a dark
energy with a time-varying density associated with a scalar field called
``quintessence'' \cite{quintessence}. As an alternative to
quintessence, Kamenshchik {\it et al.}
\cite{kamenshchik} have proposed a heuristic unification of DM and DE in terms
of an exotic fluid with an  equation of state $P=-A/\epsilon$
called the  Chaplygin gas. This equation of state provides a model of universe
that behaves as a pressureless fluid (DM) at early times, and as a
fluid with a constant energy density  (DE) at late times, yielding an
exponential acceleration similar to the effect of the cosmological constant.
However, in the intermediate regime of interest, this model does not give a good
agreement with the observations \cite{sandvik} so that various generalizations
of the Chaplygin gas model have been considered. 
In this Letter, we propose a new model based on a logotropic equation of state
\cite{logo} that seems to give a
solution to all the
problems mentioned above and, most importantly, that {\it predicts} the correct
values of $\Sigma_0$, $M_b/v_h^4$, and $M_{300}$ with remarkable accuracy, and
without free
parameter.

\section{Logotropic cosmology}
\label{sec_lc}

\subsection{The logotropic dark fluid}
\label{sec_ldf}

The Friedmann equations for a flat universe without cosmological constant
are \cite{weinbergbook}:
\begin{equation}
\frac{d\epsilon}{dt}+3\frac{\dot a}{a}(\epsilon+P)=0,\quad H^2=\left
(\frac{\dot a}{a}\right )^2=\frac{8\pi
G}{3c^2}\epsilon,
\label{l2}
\end{equation}
where $\epsilon(t)$ is the energy density, $P(t)$ is the pressure, $a(t)$ is the
scale factor, and $H=\dot a/a$ is the Hubble parameter.

For a relativistic fluid at $T=0$, or for an
adiabatic evolution (which is the
case for a perfect fluid), the first law of thermodynamics
reduces to \cite{weinbergbook}:
\begin{equation}
d\epsilon=\frac{P+\epsilon}{\rho}d\rho,
\label{l4}
\end{equation}
where $\rho$ is the rest-mass density. Combined with the equation of continuity
(\ref{l2}), we get
\begin{equation}
\frac{d\rho}{dt}+3\frac{\dot a}{a}\rho=0 \Rightarrow \rho=\frac{\rho_0}{a^3},
\label{two3}
\end{equation}
where  $\rho_0$ is the present value of the rest-mass density, and the present
value of the scale factor is taken to be $a_0=1$. This equation, which expresses
the conservation of the rest-mass, is valid for an arbitrary equation
of state.

For an equation of state specified under the form  $P=P(\rho)$, Eq.
(\ref{l4}) can be integrated to obtain the relation between the energy density
$\epsilon$ and the rest-mass density. We obtain
\begin{equation}
\epsilon=\rho c^2+\rho\int^{\rho}\frac{P(\rho')}{{\rho'}^2}\, d\rho'=\rho
c^2+u(\rho),
\label{gr7}
\end{equation}
where the constant of integration is set equal to zero. We note that $u(\rho)$
can be interpreted as an internal energy density. Therefore, the
energy density $\epsilon$ is the sum of the rest-mass energy $\rho c^2$ and the
internal energy $u(\rho)$. The rest-mass energy is positive while the internal
energy can be positive or negative. Of course, the total energy $\epsilon=\rho
c^2+u(\rho)$ is always positive.

We assume that the universe is filled with a single dark fluid described by a
logotropic equation of state
\begin{equation}
P=A\ln\left (\frac{\rho}{\rho_P}\right ),
\label{l6}
\end{equation}
where $A$ is the logotropic temperature (determined below) and
$\rho_P=5.16\times 10^{99}\, {\rm
g}\, {\rm m}^{-3}$ is the Planck
density. It will be called the Logotropic Dark Fluid
(LDF). Using Eqs. (\ref{gr7}) and (\ref{l6}), the relation
between the energy density and the rest-mass density is
\begin{equation}
\epsilon=\rho c^2-A\ln \left (\frac{\rho}{\rho_P}\right )-A=\rho c^2+u(\rho).
\label{l7}
\end{equation}
The energy density is the
sum of two
terms: a rest-mass energy term $\rho c^2\propto a^{-3}$ that mimics DM and
an internal energy term $u(\rho)=-P(\rho)-A$ that mimics DE. This decomposition
leads to a natural, and physical, unification of DM and DE and elucidates their
mysterious nature. We note that the
pressure is related to the internal energy by $P=-u-A$. Combining Eqs.
(\ref{l6}) and (\ref{l7}), we obtain  $\epsilon=\rho_P c^2 e^{P/A}-P-A$ which
determines, by inversion, the equation of state $P(\epsilon)$. From Eqs.
(\ref{two3}),
(\ref{l6}) and (\ref{l7}), we get $P=A\ln(\rho_0/\rho_Pa^3)$ and
$\epsilon=\rho_0c^2/a^3-A\ln(\rho_0/\rho_Pa^3)-A$. We note that the internal
energy
$u=-A\ln(\rho/\rho_P)-A$ is  positive for
$\rho<\rho_P/e$ and negative for $\rho>\rho_P/e$.

In the early universe ($a\rightarrow 0$, $\rho\rightarrow +\infty$),  the
rest-mass energy (DM) dominates, so that
\begin{equation}
\epsilon\sim \rho c^2\sim \frac{\rho_0 c^2}{a^3},\qquad P\sim A\ln \left
(\frac{\epsilon}{\rho_P c^2}\right ).
\label{l8}
\end{equation}
For small values of the scale factor, we recover the results of the CDM
model ($P=0$) since $\epsilon\propto a^{-3}$. In
the late universe ($a\rightarrow +\infty$, $\rho\rightarrow 0$), the internal
energy (DE) dominates, and we have
\begin{equation}
\epsilon\sim
-A\ln \left (\frac{\rho}{\rho_P}\right )\sim 3A \ln a,\qquad P\sim -\epsilon.
\label{l9}
\end{equation}
We note that the equation of state $P(\epsilon)$ behaves asymptotically  as
$P\sim -\epsilon$, similarly to the usual equation of state of DE. It is
interesting to recover the equation of state $P=-\epsilon$ from the logotropic
model (\ref{l6}). This was not obvious {\it a priori}.

\subsection{The logotropic temperature}
\label{sec_lt}

Since, in our model, the rest-mass energy of the dark fluid  mimics DM, we
identify $\rho_0$ with the present density of DM. We thus
set $\rho_0=\Omega_{m,0}\epsilon_0/c^2=2.54\times 10^{-24}\, {\rm g}\, {\rm
m}^{-3}$, where $\epsilon_0/c^2={3H_0^2}/{8\pi G}=9.26\times 10^{-24}\, {\rm
g}\, {\rm m}^{-3}$ is the present energy density of the universe (we have taken
$H_0=70.2 \,
{\rm km}\,  {\rm s}^{-1}\, {\rm Mpc}^{-1}=2.275\, 10^{-18} \,
{\rm s}^{-1}$) and 
$\Omega_{m,0}=0.274$ is the present fraction of DM (we also include baryonic
matter). As a result, the present internal energy of the dark
fluid  $u_0/c^2=\epsilon_0/c^2-\rho_0$ is identified with the present
density of DE $\rho_{\Lambda}=(1-\Omega_{m,0})\epsilon_0/c^2=6.72\times
10^{-24}\, {\rm g}\, {\rm m}^{-3}$ where
$\Omega_{\Lambda,0}=1-\Omega_{m,0}=0.726$ is the present
fraction of DE.

Applying Eq. (\ref{l7}) at $a=1$, we obtain the identity
\begin{equation}
\frac{\rho_P}{\rho_{\Lambda}}=\frac{\Omega_{m,0}}{1-\Omega_{m,0}}e^{1+1/B},
\label{l10}
\end{equation}
where we have defined the dimensionless logotropic temperature $B$ through
the relation $A=B\rho_{\Lambda}c^2$. This identity is
strikingly similar to Eq. (\ref{l1}) which appears in relation to the
cosmological constant problem. In the present context, the identity
(\ref{l10})
determines the
logotropic temperature  $B$. Qualitatively, $B\simeq
1/\ln(\rho_P/\rho_{\Lambda})\simeq 1/[123\ln(10)]$. This gives a
new interpretation to the famous number $123\simeq{\rm
log}(\rho_P/\rho_{\Lambda})$ as being the inverse logotropic
temperature. More precisely, we obtain 
\begin{equation}
B=\frac{1}{\ln\left
(\frac{1-\Omega_{m,0}}{\Omega_{m,0}}\frac{\rho_P}{\rho_{\Lambda}}\right )-1}= 
3.53\times 10^{-3}
\label{l11}
\end{equation}
and
\begin{equation}
A=B\,\rho_{\Lambda}c^2=2.13\times 10^{-9}
\, {\rm g}\, {\rm m}^{-1}\, {\rm s}^{-2}. 
\label{l12}
\end{equation}
As a result, there is no free parameter in our model. The logotropic temperature
is determined from the Planck density $\rho_P$ and the cosmological
density $\rho_{\Lambda}$  (itself obtained from the Hubble constant $H_0$ and
the fraction of  DM $\Omega_{m,0}$).
From now on,
we shall regard $A$ as a fundamental constant that supersedes
the cosmological constant. We note that it
depends on
all the fundamental constants of physics $\hbar$, $G$, $c$, and
$\Lambda$ [see Eqs. (\ref{l11}) and (\ref{l12})].

After simple manipulations, the rest-mass density, the
pressure and  the energy density of the LDF can be
expressed in terms of $B$ as 
\begin{equation}
\frac{\rho
c^2}{\epsilon_0}=\frac{\Omega_{m,0}}{a^3},\qquad \frac{P}{\epsilon_{\Lambda}}
=-B-1+B\ln\left (\frac{\rho
c^2}{\epsilon_0\Omega_{m,0}}\right ),
\end{equation}
\begin{equation}
\frac{P}{\epsilon_{\Lambda}}=-B-1-3B\ln a,
\label{pr1}
\end{equation}
\begin{equation}
\frac{\epsilon}{\epsilon_0}=\frac{\rho
c^2}{\epsilon_0}+(1-\Omega_{m,0})\left\lbrack 1+B\ln\left
(\frac{\Omega_{m,0}\epsilon_0}{\rho c^2}\right )\right\rbrack,
\label{new1}
\end{equation}
\begin{equation}
\frac{\epsilon}{\epsilon_0}=\frac{\Omega_{m,0}}{a^3}+(1-\Omega_{m,0})(1+3B\ln
a),
\label{ed9}
\end{equation}
\begin{equation}
\frac{\epsilon}{\epsilon_0}=\Omega_{m,0}e^{(B+1)/B}e^{P/B\epsilon_{\Lambda}}
-(1-\Omega_{m,0})\left (\frac{P}{\epsilon_{\Lambda}}+B\right ).
\label{pr1b}
\end{equation}

The $\Lambda$CDM model is
recovered for $B=0$, i.e., ${\epsilon}/{\epsilon_0}={\rho
c^2}/{\epsilon_0}+(1-\Omega_{m,0})$,
${\epsilon}/{\epsilon_0}={\Omega_{m ,0}}/ {a^3}+1-\Omega_{m,0}$, and 
$P=-\epsilon_{\Lambda}$. The $\Lambda$CDM model is equivalent to a
constant negative pressure
$P=-\epsilon_{\Lambda}$ \cite{cosmopoly2} and to the relation $\epsilon=\rho
c^2+\epsilon_{\Lambda}$ between the energy density and the rest-mass density.
According to Eq. (\ref{l10}), the condition $B=0$ in the
logotropic model corresponds to $\rho_P=
+\infty$, hence
$\hbar=0$. Therefore, the fact that $B$ is small but nonzero as vindicated
by the observations (see below) shows
that quantum mechanics ($\hbar\neq 0$) plays a role in the late universe in
relation to DE.

\subsection{Evolution of the logotropic universe}
\label{sec_ltw}

\begin{figure}[!ht]
\includegraphics[width=0.98\linewidth]{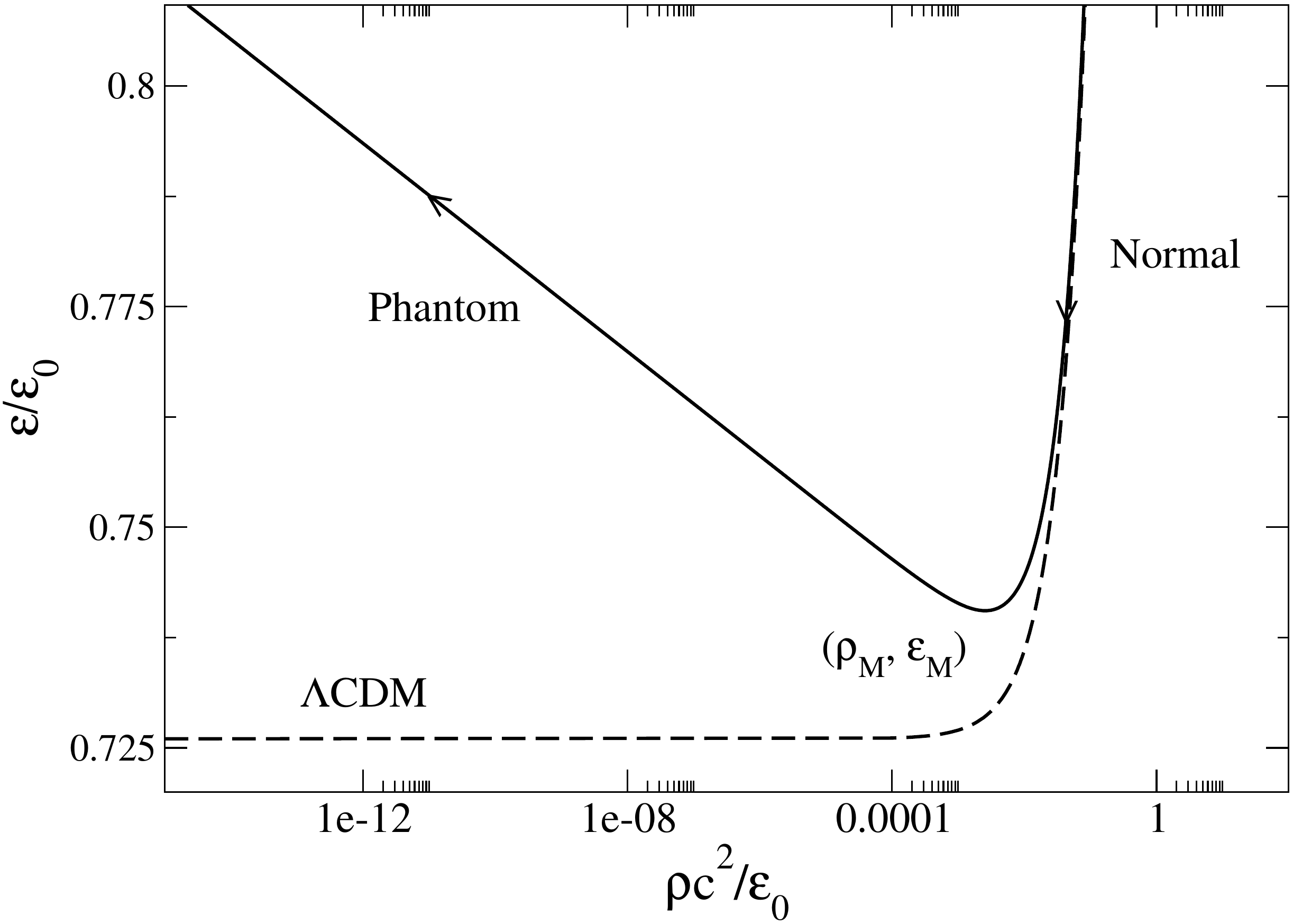}
\caption{Relation between the energy density $\epsilon$ and the rest-mass
density $\rho$ in the logotropic model. It
is compared with the relation $\epsilon=\rho c^2+\epsilon_{\Lambda}$
corresponding to the $\Lambda$CDM model. The energy density presents a
minimum $(\epsilon/\epsilon_0)_M=0.7405$ at $\rho_Mc^2/\epsilon_0=2.56\times
10^{-3}$ separating the normal universe and the phantom universe.
\label{rhoepsBpred}}
\end{figure}

\begin{figure}[!ht]
\includegraphics[width=0.98\linewidth]{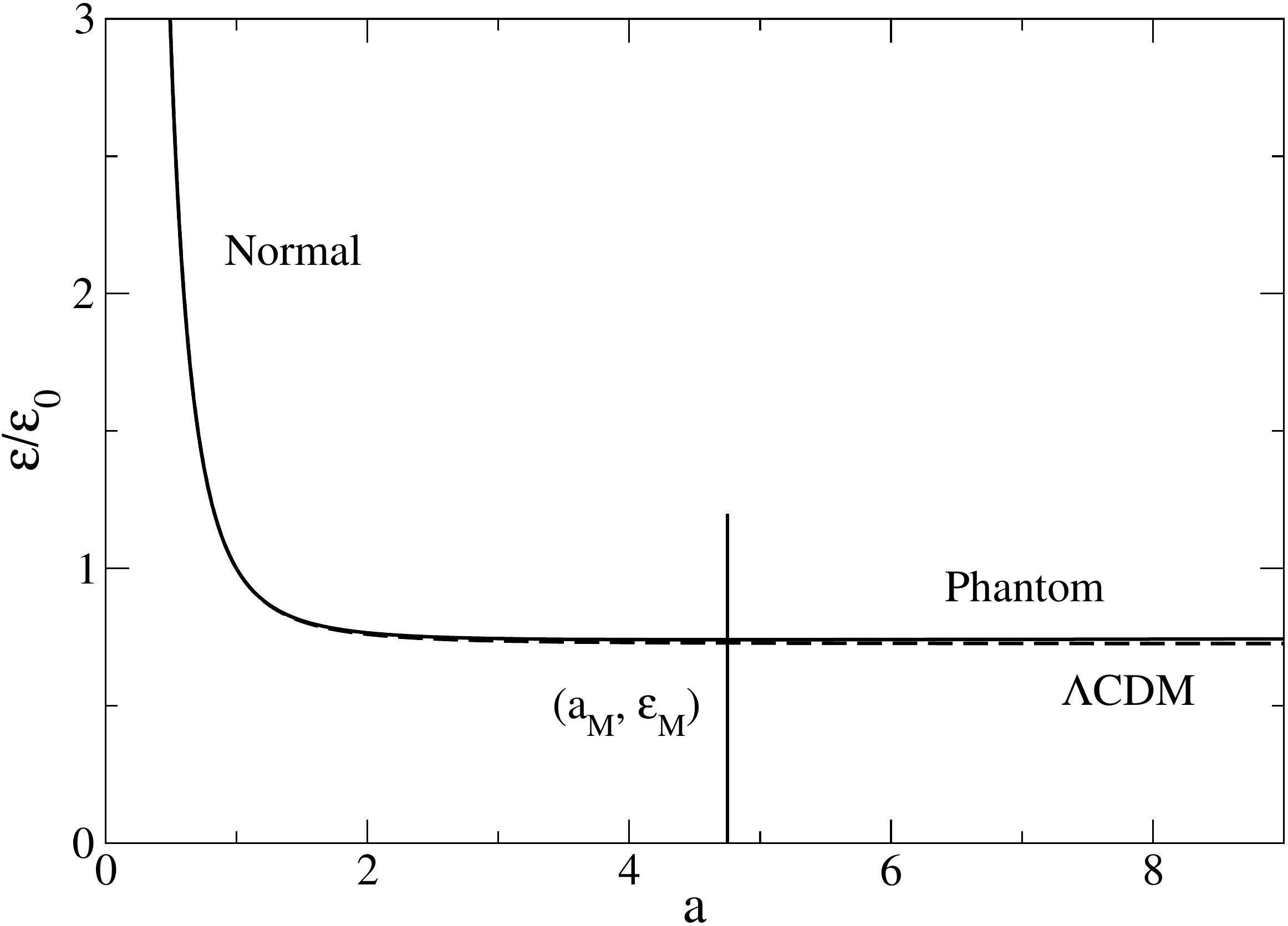}
\caption{Evolution of the energy density as a function of the scale
factor in the logotropic model. It is compared with
the $\Lambda$CDM model. The energy density presents a
minimum $(\epsilon/\epsilon_0)_M=0.7405$ at $a_M=4.75$.
\label{RepsBpred}}
\end{figure}

The relation between the energy density and the rest-mass density [see Eq.
(\ref{new1})] is plotted in
Fig. \ref{rhoepsBpred}. The evolution of the energy
density with the scale factor [see Eq.
(\ref{ed9})]  is plotted in Fig.
\ref{RepsBpred}. The universe
starts at $a=0$ with an infinite rest-mass density ($\rho\rightarrow +\infty$)
and an infinite energy density
($\epsilon\rightarrow +\infty$).\footnote{Of course, our model that attemps to
unify DM and DE is only valid at sufficiently late times, typically for
$a>a_i=10^{-4}$,
after the inflation and the radiation eras. Therefore, the limit $a\rightarrow
0$
is here formal.} The rest-mass density decreases as $a$
increases. The energy density first decreases as $a$ increases (i.e.
$\rho$
decreases), reaches a
minimum $\epsilon_M=-A\ln ({A}/{\rho_P c^2})$ at $a_M=(\rho_0c^2/A)^{1/3}$
(i.e. $\rho_M={A}/{c^2}$), then
increases as $a$ increases (i.e. $\rho$ decreases) further, and tends to
$\epsilon\rightarrow +\infty$ as $a\rightarrow +\infty$ (i.e. $\rho\rightarrow
0$). The branch $a\le a_M$ (i.e. $\rho\ge \rho_M$)  corresponds to a normal
behavior in which the energy density decreases as the scale factor increases.
The branch $a\ge a_M$ (i.e. $\rho\le \rho_M$)  corresponds to a phantom
behavior \cite{phantom} in which the energy density increases as the scale
factor increases. We
note that $A$ is equal to the rest-mass energy at the point where the universe
becomes phantom.

\begin{figure}[!ht]
\includegraphics[width=0.98\linewidth]{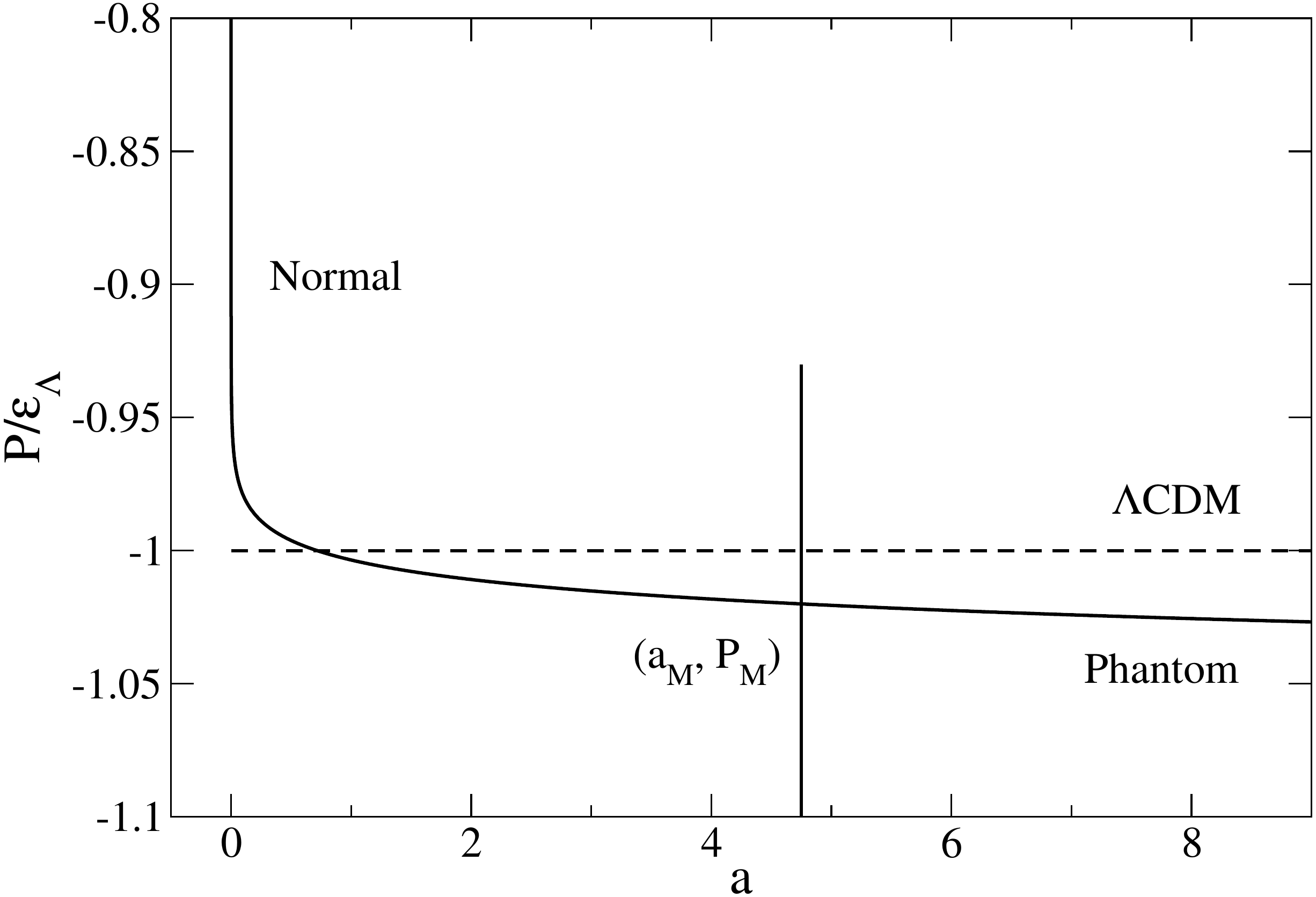}
\caption{Evolution of the pressure as a function of the scale
factor in the logotropic model. It is compared with
the $\Lambda$CDM model where $P=-\epsilon_{\Lambda}$. The pressure becomes
negative at $a_w=7.00\times
10^{-42}$ The point separating the normal universe from the phantom universe is
located at $a_M=4.75$ and
$P_M/\epsilon_{\Lambda}=-1.02$.
\label{RPBpred}}
\end{figure}

\begin{figure}[!ht]
\includegraphics[width=0.98\linewidth]{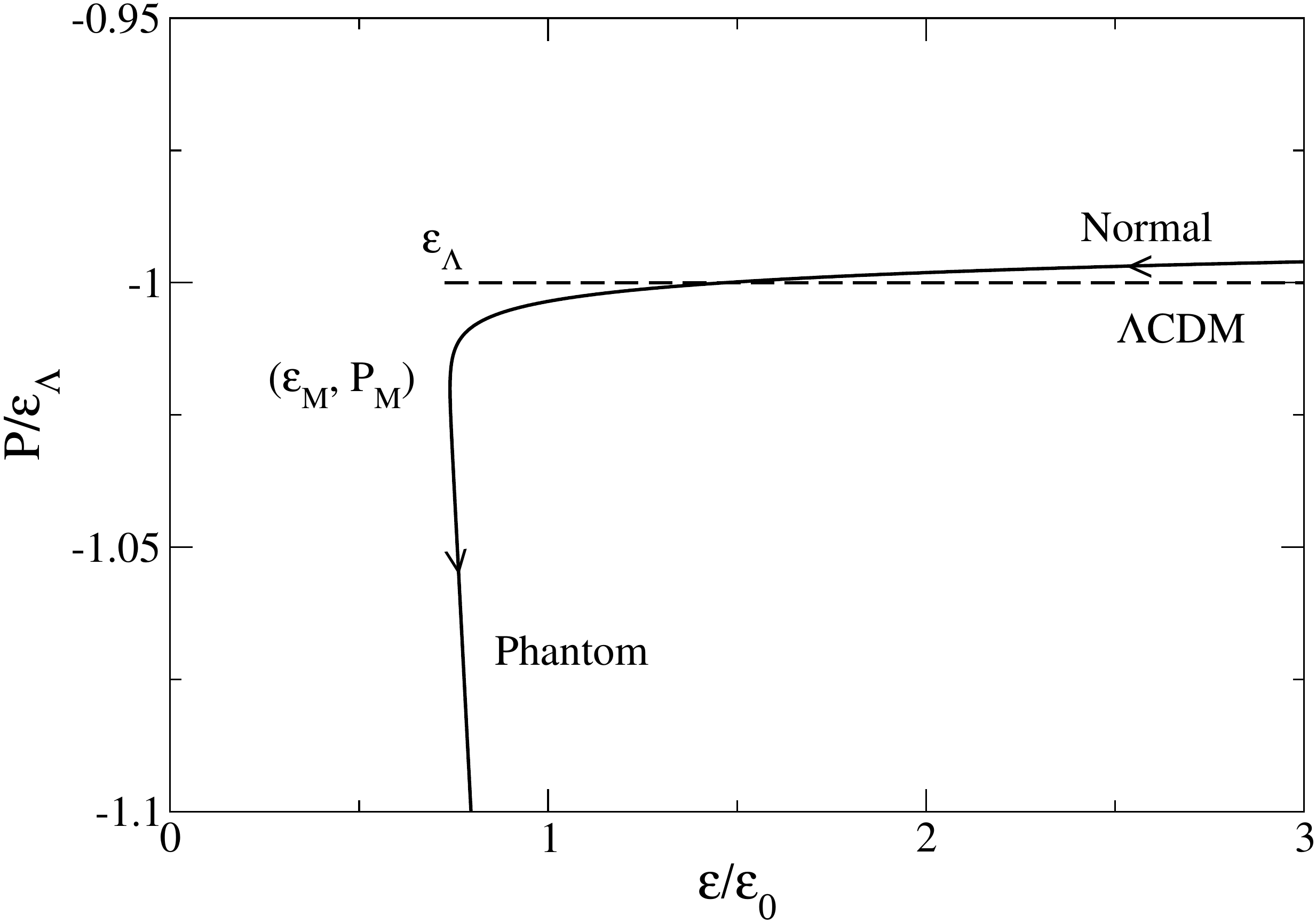}
\caption{Equation of state $P(\epsilon)$ of the logotropic model. It is
compared with the equation of state $P=-\epsilon_{\Lambda}$ of
the $\Lambda$CDM
model. 
\label{epsPBpred}}
\end{figure}

\begin{figure}[!ht]
\includegraphics[width=0.98\linewidth]{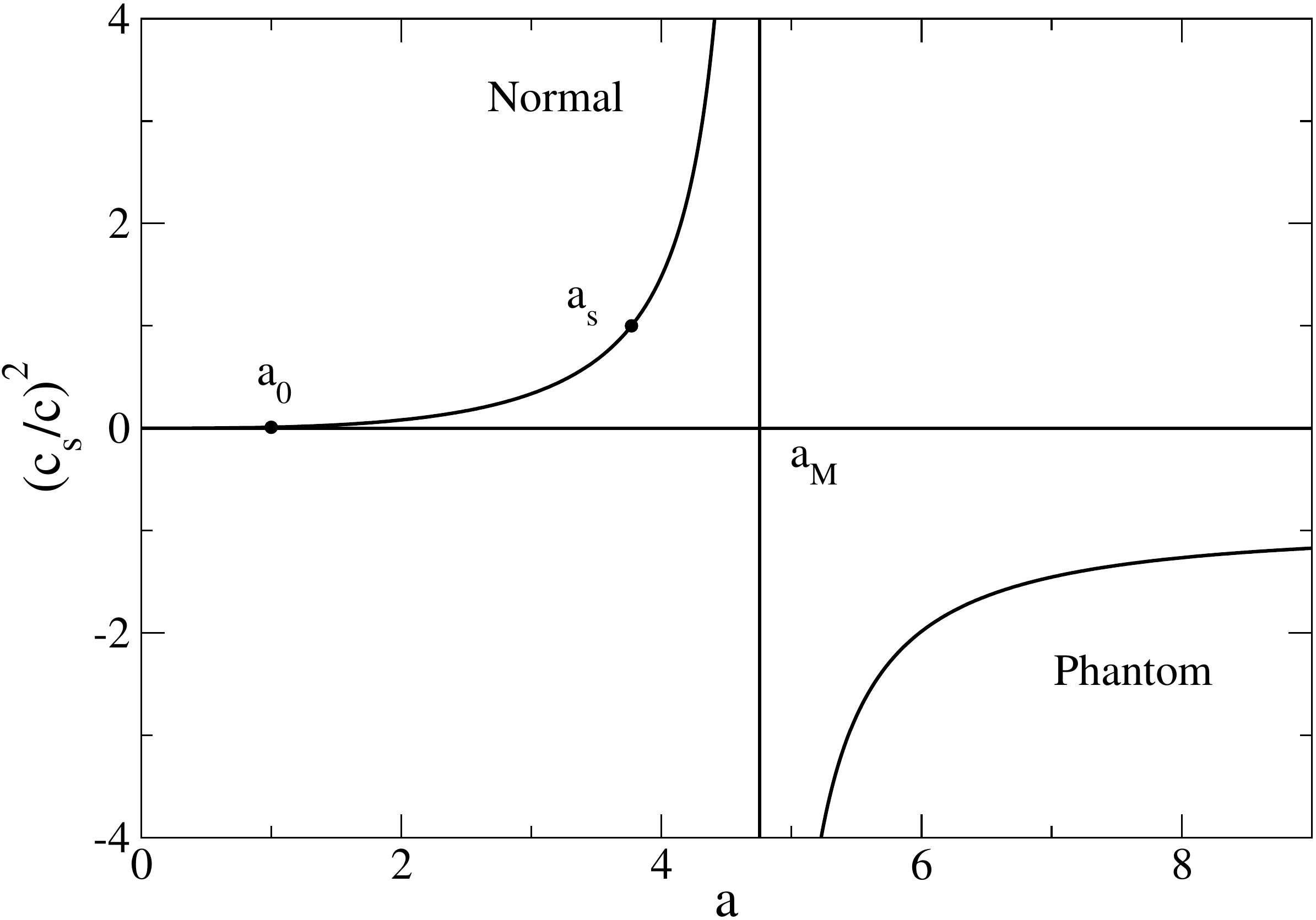}
\caption{Evolution of the speed of sound with the scale factor in
the logotropic model. For the $\Lambda$CDM
model, $c_s=0$.  The speed of sound is equal to the speed of light
($c_s=c)$ at $a_S=a_M/2^{1/3}=3.77$. At the present time ($a=a_0=1$), we have
$(c_s/c)^2=1/(a_M^3-1)=9.42\times 10^{-3}$.
\label{sound}}
\end{figure}

\begin{figure}[!ht]
\includegraphics[width=0.98\linewidth]{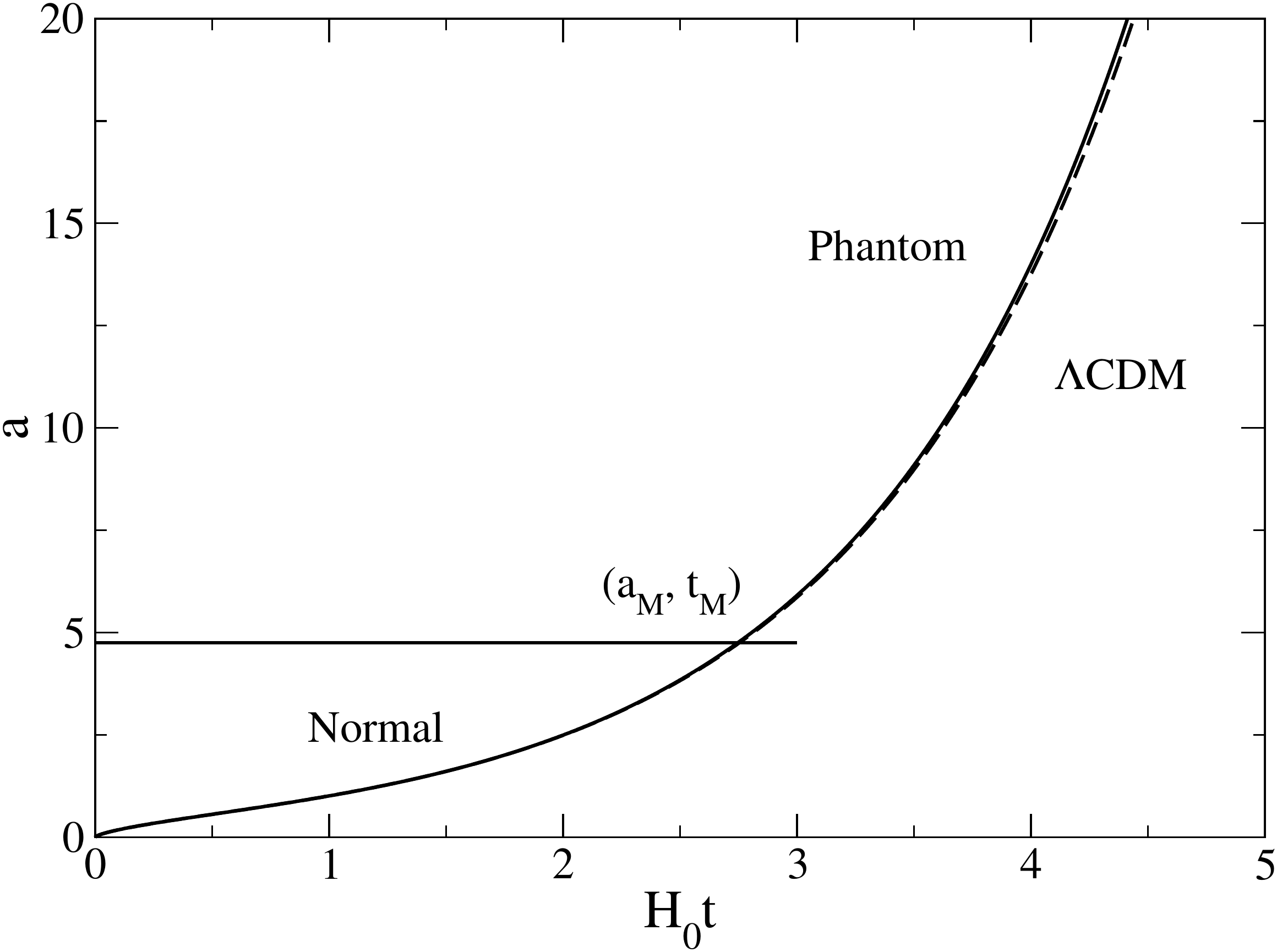}
\caption{Temporal evolution of the scale factor in
the logotropic model as compared to the $\Lambda$CDM
model. The age of the universe is $t_0=13.8\, {\rm Gyrs}$. In the logotropic
model, the universe accelerates at $t_c=7.19\, {\rm Gyrs}$, the speed of sound
exceeds the speed of light at $t_s=34.5\, {\rm
Gyrs}$, and the universe becomes phantom at $t_M=38.3\, {\rm Gyrs}$ (at present
$w_0=P_0/\epsilon_0=-0.729$).
\label{talettre}}
\end{figure}

The evolution of the pressure with the scale
factor [see Eq.
(\ref{pr1})] is plotted in Fig. \ref{RPBpred}. 
The pressure decreases as $a$ increases (i.e. $\rho$
decreases).
It starts from $P\rightarrow +\infty$ at $a=0$ (i.e. $\rho\rightarrow +\infty$,
$\epsilon\rightarrow +\infty$), vanishes at $a_w=(\rho_0/\rho_P)^{1/3}$,
achieves
the value
$P_M=-\epsilon_M$ at $a_M$ (i.e. $\rho_M$, $\epsilon_M$) and tends to
$P\rightarrow -\infty$ as
$a\rightarrow +\infty$ (i.e. $\rho\rightarrow 0$, $\epsilon\rightarrow
+\infty$). The equation of state $P(\epsilon)$ [see Eq.
(\ref{pr1b})] is defined
for $\epsilon\ge \epsilon_M$ and has two branches corresponding to a normal
universe
($P\ge P_M$) and a phantom universe ($P\le P_M$), as shown in Fig.
\ref{epsPBpred}.
Therefore,
the equation of state $P(\epsilon)$  is multi-valued.

The speed of sound $c_s$, defined by $c_s^2=P'(\epsilon)c^2$, is given by
${c_s^2}/{c^2}={1}/({\rho c^2}/{A}-1)={1}/\lbrack ({a_M}/{a})^3-1\rbrack$. It is
real for $a<a_M$ (i.e. when the
universe is
normal) and imaginary for $a>a_M$ (i.e. when the universe is phantom).
The relation between the speed of sound  and the scale factor is plotted in
Fig. \ref{sound}.

Solving the Friedmann equation (\ref{l2}) with the energy
density (\ref{ed9}), we obtain the evolution of the scale
factor represented in Fig. \ref{talettre}. For $t\rightarrow 0$, we have $a\sim
(3\sqrt{\Omega_{m,0}}H_0 t/2)^{2/3}$ and $\epsilon/\epsilon_0\sim 4/(9H_0^2t^2)$
as in the Einstein-de Sitter (EdS) model. For $t\rightarrow +\infty$, we have
$a\propto e^{3B(1-\Omega_{m,0})H_0^2t^2/4}$ and $\epsilon/\epsilon_0\sim
[3B(1-\Omega_{m,0})H_0t/2]^2$. This asymptotic solution, which is valid in the
regime
where the universe is phantom, has a super-de Sitter behavior. There is no
future finite time singularity (no Big Rip). The scale factor, the energy 
density and the pressure become infinite in infinite time (Little Rip).

\subsection{Cosmological implications}
\label{sec_ci}

From the
observational
viewpoint, there is
no visible difference between the logotropic model and the $\Lambda$CDM model at
large scales.
Differences will appear in about $25\, {\rm Gyrs}$, when the universe
becomes phantom (this aspect will be developed in a future work).
However, this moment is very remote in the future, and for the time being, the
logotropic model and the  $\Lambda$CDM model behave similarly (see Figs.
\ref{RepsBpred} and \ref{talettre}). This is
satisfactory since the $\Lambda$CDM model works well at the cosmological
scale. On a theoretical point of view, the logotropic model has several
advantages with respect to the  $\Lambda$CDM model.
In our model, DM and DE are the manifestation of a single dark fluid described
by a unique equation of state. Therefore, there is no cosmic coincidence
problem. On the other hand, the cosmological constant
problem of Eq. (\ref{l1}) is translated into an equation (\ref{l10}) that
determines the logotropic temperature $A\simeq\epsilon_{\Lambda}/[123\ln(10)]$.

An important difference between the $\Lambda$CDM model and the logotropic model
concerns the speed of sound $c_s$ defined by
$c_s^2/c^2=P'(\epsilon)$. In the
$\Lambda$CDM model, since $P=0$ (actually, $P=-\epsilon_{\Lambda}$), the
speed of sound $c_s=0$. As a result, the
Jeans length is zero ($\lambda_J=0$), implying that the homogeneous background
is unstable at all scales so that halos of any size should be observed in
principle. However, this is not the case. There does not seem to be halos with a
size
smaller than $R_{\rm min}\sim 10\, {\rm pc}$. Contrary to the $\Lambda$CDM
model, the
logotropic model has a nonzero speed of sound, hence a nonzero Jeans length.
We
can obtain an estimate of the Jeans length $\lambda_J$ at
the beginning of the matter era where perturbations start to grow. We assume
that the matter era starts at $a_i=10^{-4}$, corresponding to the epoch of
matter-radiation equality. In this era, we can make the approximation
$\epsilon=\rho c^2$, so the Jeans wavenumber is given by 
$k_J^2={4\pi G\rho a^2}/{c_s^2}$ \cite{weinbergbook}, where
$c_s^2=P'(\rho)=A/\rho$. At
$a_i=10^{-4}$, we find $\rho_i=2.54\times
10^{-12} {\rm g}/{\rm m}^3$ and $(c_s^2/c^2)_i=9.33\times 10^{-15}$. This leads
to a Jeans length $\lambda_J=2\pi/k_J=1.25\times 10^{18}\, {\rm m}=40.4\, {\rm
pc}$ which is of the order of magnitude of the smallest known dark matter halos
such as Willman I ($r_h=33\, {\rm pc}$) (see Table 2 of \cite{destri}). We
predict that there should not exist
halos of smaller size since the perturbations are stable for
$\lambda<\lambda_J$. This is in agreement with the observations, unlike the
$\Lambda$CDM model. Therefore, a small but finite value of $B$, yielding a
nonzero speed of sound and a nonzero Jeans length, is able to account for
the minimum observed size of dark matter halos in the universe. It also puts a
cut-off in the density power spectrum of the perturbations and sharply
suppresses small-scale linear power. This may be a way to solve the missing
satellite problem.

\section{Logotropic dark matter halos}
\label{sec_ldm}

The interest of the logotropic model becomes evident when it is
applied to DM halos. We assume that DM halos are
described by the logotropic equation of state (\ref{l6}) with the
logotropic temperature $A=2.13\times 10^{-9} \, {\rm g}\, {\rm m}^{-1}\, {\rm
s}^{-2}$ determined previously, viewed as a
fundamental constant. At the galactic scale, we can use Newtonian gravity.
Combining the  condition of hydrostatic equilibrium $\nabla
P+\rho\nabla\Phi={\bf 0}$ with the Poisson equation
$\Delta\Phi=4\pi G\rho$, assuming spherical symmetry, and introducing the
notations $\theta={\rho_0}/{\rho}$ and $\xi={r}/{r_0}$, where $\rho_0$ is the
central density and 
\begin{equation}
r_0=\left (\frac{A}{4\pi G\rho_0^2}\right )^{1/2}
\label{l14}
\end{equation}
is the logotropic core radius, we obtain the differential equation
\begin{equation}
\frac{1}{\xi^2}\frac{d}{d\xi}\left (\xi^2\frac{d\theta}{d\xi}\right
)=\frac{1}{\theta}
\label{l15}
\end{equation}
with $\theta(0)=1$ and $\theta'(0)=0$. It can be viewed as a Lane-Emden
equation of index $n=-1$ \cite{chandra}. This
equation has a simple analytical solution  $\rho_s=(A/8\pi
G)^{1/2} r^{-1}$ called the singular logotropic sphere because it diverges at
the origin \cite{logo}.\footnote{We note, parenthetically, that this singular
solution $\propto r^{-1}$ is similar to NFW cusps \cite{nfw}.} The regular
solutions must be computed numerically.
They have a flat core and behave as  $\rho\sim (A/8\pi
G)^{1/2} r^{-1}$ for $r\rightarrow +\infty$. Since the logotropic spheres are
homologous, they generate a 
{\it universal} DM profile. Indeed, if we rescale the density by the central
density $\rho_0$ and the radial distance by the core radius $r_0$, we get an
invariant density profile $1/\theta(\xi)$. We note that the total mass of a
logotropic sphere is infinite because of the slow decay of the density. This
means that the logotropic distribution
cannot describe the whole cluster. It is valid only in the core. At larger
distances, we must take into account complex physical processes such as tidal
effects and incomplete relaxation that steepen the density profile (see, e.g.,
\cite{clm}). It is usually found that the density profiles of DM halos decrease
at large distances as $r^{-3}$ \cite{nfw,observations}.
Since we do not take these complicated processes
into account, our logotropic model is only valid up to a few values of
the core radius $r_0$. However, this is sufficient to determine the physical
characteristics of
DM halos.

\begin{figure}[!ht]
\includegraphics[width=0.98\linewidth]{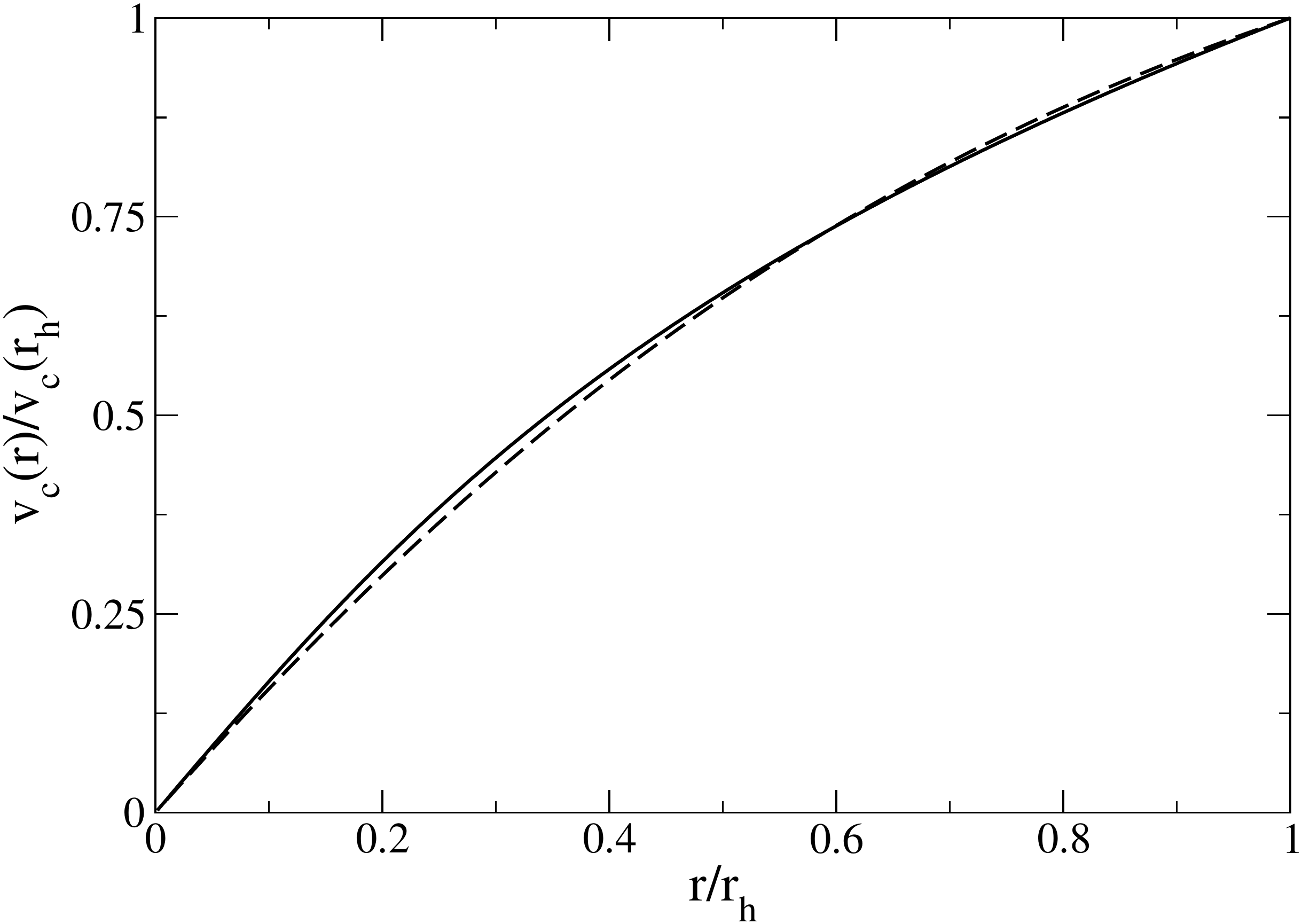}
\caption{Circular velocity profile of a logotropic sphere (solid line)
in the region $r\le r_h$ where the model is valid. It is compared with the
Burkert profile (dashed line).
\label{vitLOGOzoom}}
\end{figure}

Using the Lane-Emden equation (\ref{l15}) the mass
profile $M(r)=\int_0^r
\rho(r')\, 4\pi {r'}^2\, dr'$ is given by $M(r)=4\pi \rho_0 r_0^3 \xi^2
\theta'(\xi)$. The circular velocity defined by $v_c^2(r)={GM(r)}/{r}$ can be
expressed as $v_c^2(r)=4\pi G \rho_0 r_0^2 \xi \theta'(\xi)$.
We define the halo radius $r_h$ as the radius at which
$\rho/\rho_0=1/4$. The dimensionless halo radius is the solution of the equation
$\theta(\xi_h)=4$. We numerically find $\xi_h=5.8458$ and
$\theta'(\xi_h)=0.69343$.
Then, $r_h=\xi_h r_0$. The normalized halo mass at the halo radius is given by
\begin{equation}
\frac{M_h}{\rho_0 r_h^3}=4\pi
\frac{\theta'(\xi_h)}{\xi_h}=1.49.
\label{l17}
\end{equation}
This value is  relatively close to the value ${M_h}/{\rho_0 r_h^3}=1.60$
\cite{vega,clm} obtained from the empirical Burkert profile 
\cite{observations} 
that provides a good fit of DM halos. On the other hand, the
universal rotation curve predicted by the logotropic model is very close to the
Burkert profile  up to the halo radius, i.e. for $r\le r_h$ (see Fig.
\ref{vitLOGOzoom}). Very recently, Burkert \cite{burkertnew} observed that the
density
profile of real DM halos behaves approximately as $r^{-1}$ close to the halo
radius. Interestingly, we note that the exponent $-1$ precisely corresponds to
the characteristic exponent  of the logotropes.

In addition to these already encouraging results, the logotropic equation of
state has a very interesting property. According to Eq. (\ref{l14}), the
surface density of the logotropic sphere is given by
\begin{equation}
\Sigma_0\equiv \rho_0 r_h=\left (\frac{A}{4\pi G}\right )^{1/2}\xi_h.
\label{l18}
\end{equation}
Since the logotropic temperature $A$ is the same for all the halos (a
consequence of our approach where we view $A$  as a fundamental constant), this
implies
that the surface density of the DM halos should be the
same. This is precisely what is observed \cite{donato}. Using the
value of the logotropic temperature given by Eq. (\ref{l12}), we get
$\Sigma_0^{\rm th}=141\, M_{\odot}/{\rm pc}^2$  which coincides with the
best-fit value $\Sigma_0^{\rm obs}=141_{-52}^{+83}\, M_{\odot}/{\rm pc}^2$ of
the surface density of DM halos \cite{donato}. This agreement is
remarkable since there is no free parameter in our model. Furthermore, it is
non trivial since the constant $A$ depends, through Eqs. (\ref{l11}) and
(\ref{l12}), on the Planck density $\rho_P$ and on the cosmological density
$\rho_{\Lambda}$. This suggests that there is something deep behind these
relations.

There are interesting consequences of this result.
According to Eq. (\ref{l17}), the mass of the halos calculated at the halo
radius $r_h$ is given by  $M_h=1.49\Sigma_0 r_h^2$. On the other hand, the
circular
velocity at the halo radius is $v_h^2=GM_h/r_h=1.49\Sigma_0
G r_h$. Since the surface density of
the dark matter halos is constant, we obtain
$M_h/M_{\odot}=210 (r_h/{\rm pc})^2\propto r_h^2$ and  $(v_h/{\rm
km}\, {\rm s}^{-1})^2=0.905(r_h/{\rm pc})\propto r_h$. These scalings are
consistent with the observations \cite{vega}. Furthermore, introducing the
baryon
mass $M_b=f_b M_h$ where
$f_b\sim 0.17$ is the cosmic baryon fraction \cite{mcgaugh}, we get
\begin{equation}
\frac{M_b}{v_h^4}=\frac{f_b}{1.49\Sigma_0 G^2}=\frac{f_b}{\theta'(\xi_h)(4\pi
AG^3)^{1/2}}.
\label{tf}
\end{equation}
Therefore, $v_h^4\propto M_h$ which is the TF relation. More
precisely, we predict
$(M_b/v_h^4)^{\rm th}=44 M_{\odot}{\rm km}^{-4}{\rm s}^4$ which is close to the
observed value 
$(M_b/v_h^4)^{\rm obs}=47\pm 6 M_{\odot}{\rm km}^{-4}{\rm s}^4$
\cite{mcgaugh} (we obtain a perfect agreement by taking
$f_b=0.18$).\footnote{The TF relation is sometimes justified by
the MOND theory \cite{mond} which predicts a relation of the form $v_h^4=Ga_0
M_b$ between the asymptotic circular velocity and the baryon mass, where $a_0$
is a critical acceleration. Our results imply
$a_0^{\rm th}=1.72\times
10^{-10}\, {\rm m}/{\rm s}^2$ which is
close to the value $a_0^{\rm obs}=(1.3\pm 0.3)\times
10^{-10}\, {\rm m}/{\rm s}^2$ obtained from the observations \cite{mcgaugh}.
Using the results of the Appendix, we have $a_0^{\rm th}\simeq H_0 c/4$ which
explains why $a_0$ is of the order $H_0 c$. We emphasize, however, that we do
not use the MOND theory in our approach.}

The logotropic equation of state also explains the observation of Strigari {\it
et al.} \cite{strigari} that all dSphs of the Milky
Way have the same total dark matter mass $M_{300}$ contained within a radius
$r_u=300\, {\rm pc}$, namely  $\log
(M_{300}^{\rm obs}/M_{\odot})=7.0^{+0.3}_{-0.4}$.
Using $M_{300}=4\pi\rho_0r_0^3\xi_u^2\theta'(\xi_u)$, $\xi_u=r_u/r_0$ and
$r_0=r_h/\xi_h$, we obtain $M_{300}=4\pi \rho_0 r_h ({r_u^2}/{\xi_h})
\theta'\left (\xi_h {r_u}/{r_h}\right )$. The logotropic distribution has the
asymptotic behavior $\theta(\xi)\sim
\xi/\sqrt{2}$ for $\xi\rightarrow +\infty$ \cite{logo}. For the dSphs
considered in
\cite{strigari}, $\xi_h r_u/r_h\gg 1$ (see Table 2 of \cite{destri}) so
$\theta'(\xi_h {r_u}/{r_h})$ can be
replaced
by its asymptotic value $1/\sqrt{2}$. This yields
\begin{equation}
M_{300}= \frac{4\pi \Sigma_0 r_u^2}{\xi_h\sqrt{2}}=r_u^2 \left (\frac{2\pi
A}{G}\right )^{1/2},
\label{l20}
\end{equation}
which is a constant, in agreement with the
claim of Strigari {\it et al.}
\cite{strigari}. We note that the  constancy of $M_{300}$ is due to the
universality of $A$. Furthermore, the numerical application gives
$M_{300}^{\rm th}=1.93\times 10^{7}\, M_{\odot}$, leading to $\log
(M_{300}^{\rm th}/M_{\odot})=7.28$ in very good agreement with the
observational value.

In conclusion, the logotropic equation of state can simultaneously account for
cosmological constraints (with the same level of precision as the $\Lambda$CDM
model) and explain properties of DM halos (their minimum size $R_{\rm min}$,
their surface density $\Sigma_0$, their mass $M_{300}$, and the TF ratio
$M_b/v_h^4$) that were not
explained
so
far. This may be a hint that DM and DE are the manifestations
of a single dark fluid. The best illustration of this ``unification'' is that
we have obtained the value of $A$ from cosmological constraints [see Eqs.
(\ref{l11}) and (\ref{l12})], and that this value accounts for the universality
of the surface density
$\Sigma_0$ and mass $M_{300}$ of DM halos [see Eqs. (\ref{l18}) and
(\ref{l20})], as well as for the TF relation [see Eq. (\ref{tf})]. Assuming that
this agreement is not a coincidence (the
perfect agreement between the predicted values of $\Sigma_0$, $M_{300}$,
$M_b/v_h^4$ and
the observations is a strong support to our approach), the next step is
to justify the logotropic equation of state. We sketch below several possible
justifications based on extra-dimensions (Cardassian model), generalized
thermodynamics, and field theory.

\section{Possible justifications of the logotropic equation of state}
\label{sec_just}

\subsection{Cardassian model}
\label{sec_card}

Freese and Lewis \cite{freese}, in their so-called Cardassian
model, have
proposed to explain the accelerated expansion of the universe in terms of a
modified Friedmann equation of the form
\begin{equation}
H^2=\frac{8\pi
G}{3}\rho+\nu(\rho),
\label{l21}
\end{equation}
where $\rho=\rho_0 a^{-3}$ is the rest-mass density and $\nu(\rho)$ is a
``new'' term which characterizes the model. In the early universe, the
term $\nu(\rho)$ is negligible and one recovers the usual Friedmann equation of
pressureless matter leading to a decelerating universe with $a\propto
t^{2/3}$ (Einstein-de Sitter solution). In the late universe, the term
$\nu(\rho)$ dominates and causes the universe to accelerate. Freese and Lewis
\cite{freese} justify the modified Friedmann equation (\ref{l21}) in relation to
the existence of extra-dimensions. The usual Friedmann equation is modified as a
consequence of embedding our
universe as a three-dimensional surface ($3$-brane) in higher dimensions.
Our approach provides another, simpler, justification of this equation from the
ordinary four dimensional Einstein
equations. Starting from the usual Friedmann equation
(\ref{l2}) and considering a dark fluid at $T=0$, or an adiabatic fluid, with
the energy density given by $\epsilon=\rho c^2+u(\rho)$ [see Eq.
(\ref{gr7})], we
obtain Eq. (\ref{l21}) with $\nu(\rho)=({8\pi G}/{3c^2}) u(\rho)$.
Therefore, the
``new'' term in the ``modified'' Friedmann
equation (\ref{l21}) can be
interpreted as the internal
energy $u(\rho)$ of the dark fluid while
the ``ordinary'' term $8\pi
G\rho/3$ corresponds to its rest-mass energy density $\rho c^2$.
Therefore, our approach provides a new justification of the Cardassian model.
Inversely, the original justification of the Cardassian model, namely that the
term $\nu(\rho)$ 
arising in the modified Friedmann equation (\ref{l21}) may result from the
existence of extra-dimensions, could be a way to justify the logotropic
model corresponding to $\nu(\rho)=-({8\pi GA}/{3c^2})\left\lbrack \ln\left
({\rho}/{\rho_P}\right )+1\right\rbrack$.
In this
respect,
we note that the logotropic model
asymptotically yields an equation of state of the form  $P\sim-\epsilon$ [see
Eq. (\ref{l9})]. Using the virial theorem, one can easily show 
that this equation
of state arises from a long-range confining force  $F_{ij}=-3U_0 r_{ij}^2$
that could be a fifth force or an effective
description of higher dimensional physics \cite{freese}.

\subsection{Generalized thermodynamics}
\label{sec_gt}

The logotropic equation of state was introduced phenomenologically in
astrophysics
by McLaughlin and  Pudritz \cite{pud} to describe the internal
structure and the
average properties of molecular clouds and clumps. It was also studied by
Chavanis and Sire \cite{logo} in the context of Tsallis generalized
thermodynamics \cite{tsallisbook} where it was shown to
correspond to a
polytropic equation of state of the form $P=K\rho^{\gamma}$ with
$\gamma\rightarrow 0$ and $K\rightarrow \infty$ in such a way
that $A=\gamma K$ is finite. It is associated with a generalized entropy of the
form 
\begin{equation}
S_{L}=\int \ln\rho\, d{\bf r},
\label{l22}
\end{equation}
which is called the Log-entropy \cite{logo}. The free energy can be written as
$F_L=E-A S_{L}$, where $E=(1/2)\int\rho\Phi\, d{\bf r}$
is the gravitational energy. A critical point of $F_L$ at fixed mass
$M=\int\rho\, d{\bf r}$, determined by the Euler-Lagrange equation
$\delta F_L-\mu\delta M=0$,
where $\mu$ is a Lagrange multiplier (chemical potential), leads to the
Lorentzian-type distribution $\rho({\bf r})={1}/[{\alpha+\Phi({\bf
r})/A}]$, where $\alpha=-\mu/A$. We can check that this equation is equivalent
to the
equation of hydrostatic
equilibrium with the logotropic equation of state (\ref{l6}). When
combined with the Poisson
equation, we recover the Lane-Emden equation
(\ref{l15}).  These considerations show that $A$
can be interpreted as a generalized
temperature. This is why we  call it the logotropic temperature.  As a result,
the universality of $A$ (which explains the constant values of $\Sigma_0$,
$M_{300}$ and $M_b/v_h^4$) may be
interpreted by saying that the universe is ``isothermal'', except
that isothermality does not refer to a linear equation of state but to a
logotropic equation of state in a generalized thermodynamical framework. 
If our model is correct, it would be a nice confirmation of the
interest of generalized thermodynamics \cite{tsallisbook} in physics and
astrophysics.

\subsection{Scalar field theory}
\label{sec_sf}

The logotropic equation of state can also be justified from a scalar field
theory. If we view the dark fluid as a scalar field representing BECs, its
evolution is described, in the nonrelativistic regime
appropriate to DM halos, by the Gross-Pitaevskii equation
\cite{prd1}. The
GP equation associated with the  logotropic equation of state takes the form
\begin{equation}
i\hbar \frac{\partial\psi}{\partial
t}=-\frac{\hbar^2}{2m}\Delta\psi+m\Phi\psi-Am\frac{1}{|\psi|^2}\psi,
\label{l23}
\end{equation}
where $\psi({\bf r},t)$ is the wave function. It is coupled to the
Poisson equation $\Delta\Phi=4\pi G|\psi|^2$. Eq. (\ref{l23}) can be viewed as a
GP equation with an inverted quadratic
potential, i.e. with the exponent $-2$ instead of $+2$ in the usual GP equation
\cite{revuebec}. To our knowledge, this equation has not been
introduced before. Using the Madelung
\cite{madelung} transformation, this equation can be
written in the form of fluid
equations, called the quantum Euler equations \cite{prd1}, incorporating a
quantum potential $Q=-({\hbar^2}/{2m}){\Delta\sqrt{\rho}}/{\sqrt{\rho}}$
due to the kinetic term $-({\hbar^2}/{2m})\Delta\psi$ (Heisenberg) and an
isotropic pressure
$P$  due
to the interaction term $Am|\psi|^{-2}\psi$. For the inverted quadratic
potential,  the pressure $P$ is given by the logotropic equation of state
(\ref{l6}). In the Thomas-Fermi approximation, one can
neglect the quantum potential. In that case, we recover the classical
Euler-Poisson  equations. A steady state of these equations satisfies the
condition of hydrostatic equilibrium. Furthermore, one can show
that  the free energy $F_L$ and the mass $M$ are conserved by the
Euler-Poisson equations. As a result, a minimum of free energy $F_L$ at fixed
mass $M$ is a stable steady state of the Euler-Poisson
equations  \cite{prd1}. This makes a
correspondance with
generalized thermodynamics and shows the self-consistency of our approach. We
also note that the GPP equations can be obtained as the nonrelativistic limit
$c\rightarrow +\infty$ of the Klein-Gordon-Einstein equations \cite{abrilph}.
The KG equation
corresponding to the logotrope has a logarithmic potential
$V(|\phi|)=-2A\ln|\phi|$ and writes
\begin{eqnarray}
\frac{1}{c^2}\frac{\partial^2\phi}{\partial
t^2}-\Delta\phi+\frac{m^2c^2}{\hbar^2}\left (1+\frac{2\Phi}{c^2}\right
)\phi-\frac{2A}{|\phi|^2}\phi=0.
\end{eqnarray}

\section{Conclusion}
\label{sec_conclusion}

We have proposed a heuristic unification of DM and DE in terms of a
single dark fluid with a logotropic equation of state (LDF). According to our
model, what we
usually call DM corresponds to the rest-mass density of the dark fluid and what
we
usually call DE corresponds to the internal energy of the dark fluid.

At the cosmological scale, our model satisfies the same observational
constraints as the $\Lambda$CDM model but avoids the cosmic coincidence problem
(since DM and DE are the manifestation of a single dark fluid) and the
cosmological constant problem (since there is no cosmological constant in our
approach). It also has a nonzero speed of sound and a nonzero Jeans length
(contrary to the $\Lambda$CDM model) which can explain the minimum size
 $R_{\rm min}\sim 10\, {\rm pc}$ of DM halos. 

At the galactic scale, the logotropic pressure balances gravitational
attraction and solves the cusp problem and the missing satellite problem of the
CDM model. On the other hand, the logotropic model generates a universal
rotation curve that provides a good agreement with the Burkert profile up to the
halo radius. Furthermore, it implies that the surface density
of DM halos and the mass of dwarf halos are the same for all the halos, in
agreement with the observations. It also implies the TF relation.

The most striking property of the logotropic model is the following. Using
cosmological observations, we can obtain the
value of the logotropic temperature $A=2.13\times 10^{-9} \, {\rm g}\, {\rm
m}^{-1}\, {\rm s}^{-2}$ [see Eq. (\ref{l12})]. It may be viewed
as a fundamental constant  since it actually depends on all the
fundamental constants of physics $\hbar$, $G$, $c$, and $\Lambda$. Then,
applying the
logotropic model to DM halos, and using this value of $A$, we can obtain the
value of $\Sigma_0$ [see Eq. (\ref{l18})], $M_b/v_h^4$ [see Eq.
(\ref{tf})] and $M_{300}$ [see Eq.
(\ref{l20})]  which
are in perfect agreement with the observations. Therefore,
the logotropic model is able to account both for cosmological and galactic
observations remarkably well. This may be a hint that DM
and DE are the manifestation of a unique dark fluid.

Finally, we have sketched some possible justifications of the logotropic
equation of state in relation to the existence of extra-dimensions (Cardassian
model), in relation to Tsallis generalized thermodynamics, and in relation
to scalar field theory and BECs.

The fact that the Planck density $\rho_P$ enters in the
logotropic equation of
state (\ref{l6}) designed to model DM and DE is intriguing.
It suggests that quantum mechanics manifests itself at
the cosmological scale in relation to DE. This may be a hint for a 
fundamental theory of quantum gravity. This also suggests that the logotropic
equation of state may be the limit of a more general equation of state
providing a possible
unification of DE ($\rho_{\Lambda}$) in the late universe and inflation 
(vacuum energy $\rho_{P}$) in the primordial universe. These open questions are
a strong incentive to study the logotropic model further in future works.
The phantom properties of the logotropic model will be discussed in a specific
paper.

\appendix

\section{Expression of the observational quantities in terms of the fundamental
constants}

We enlight the remarkable feature that, in our theory, all the observational
quantities can be predicted  in terms of  fundamental constants such as
$\hbar$, $G$, $c$, $H_0$, and $\Omega_{m,0}$.
We have $B={1}/\lbrack\ln\left ({8\pi c^5}/{3\Omega_{m,0}\hbar G
H_0^2}\right )-1\rbrack$. We introduce the notation $\chi=\left\lbrack
{3}B(1-\Omega_{m,0})/2\right\rbrack^{1/2}=6.20\times
10^{-2}$. Then $A={\chi^2c^2H_0^2}/{4\pi G}=3.06\times 10^{-4}
{c^2H_0^2}/{G}$, $\Sigma_0=\chi\xi_h{
H_0c}/{4\pi G}=2.89\times 10^{-2}
{H_0c}/{G}$, ${M_{h}}/{r_h^2}=\chi\theta'(\xi_h){H_0c}/{G}=4.30\times
10^{-2} {H_0c}/{G}$, ${v_{h}^2}/{r_h}=\chi\theta'(\xi_h)H_0c=4.30\times
10^{-2} H_0c$, ${v_{h}^4}/{M_b}=\chi\theta'(\xi_h){G
H_0c}/{f_b}=4.30\times 10^{-2}{GH_0c}/{f_b}$, $a_0=\chi\theta'(\xi_h)
{H_0c}/{f_b}=4.30\times 10^{-2}{H_0c}/{f_b}$,
and $M_{300}/r_u^2=\chi{H_0c}/{\sqrt{2}G}
=4.39\times 10^{-2}{H_0c}/{G}$. Noting that $\epsilon_{\Lambda}=\Lambda
c^2/8\pi G$, $\epsilon_{\Lambda}=(1-\Omega_{m,0})\epsilon_0$ and $H_0^2=(8\pi
G/3c^2)\epsilon_0$, we obtain
$\Lambda=3(1-\Omega_{m,0})H_0^2=2\chi^2H_0^2/B=1.13\times 10^{-35}\, {\rm
s}^{-2}$. Therefore, the observational quantities can be expressed equivalently
in terms of
$\hbar$, $G$, $c$, $\Lambda$ and $\Omega_{m,0}$.

\end{document}